\documentclass[prc,twocolumn,superscriptaddress,showpacs,preprintnumbers,amsmath,amssymb]{revtex4}
\usepackage[dvips]{graphicx}
\usepackage{dcolumn}
\usepackage{bm}
\usepackage{amsmath,amssymb}
\usepackage{longtable}
\usepackage{tabularx}
\def\m@thcombine#1#2{%
  \setbox0=\hbox{$#1$}
  \setbox1=\hbox{$#2$}
  \ifdim\wd0>\wd1
    \setbox0=\hbox to\wd1{\hss\box0\hss}
  \else
    \setbox1=\hbox to\wd0{\hss\box1\hss}
  \fi
  \mathop{\vcenter{
    \offinterlineskip\box0\box1}}}
\def\lesim{\m@thcombine<\sim}
\def\gesim{\m@thcombine>\sim}

\newcommand{\bra}[1]{\langle {#1} |}
\newcommand{\ket}[1]{| {#1} \rangle}

\begin{document}
\title{Self-consistent calculation of
nuclear photoabsorption cross section:
Finite amplitude method with Skyrme functionals in the three-dimensional
real space}
\author{Tsunenori Inakura}
\affiliation{Center for Computational Sciences, University of Tsukuba, Tsukuba 305-8571, Japan}
\affiliation{Theoretical Nuclear Physics Laboratory, RIKEN Nishina Center, Wako, 351-0198, Japan}
\author{Takashi Nakatsukasa}
\affiliation{Theoretical Nuclear Physics Laboratory, RIKEN Nishina Center, Wako, 351-0198, Japan}
\affiliation{Center for Computational Sciences, University of Tsukuba, Tsukuba 305-8571, Japan}
\author{Kazuhiro Yabana}
\affiliation{Center for Computational Sciences, University of Tsukuba, Tsukuba 305-8571, Japan}
\affiliation{Theoretical Nuclear Physics Laboratory, RIKEN Nishina Center, Wako, 351-0198, Japan}

\begin{abstract}
The finite amplitude method (FAM), which we have recently 
proposed (T. Nakatsukasa, T. Inakura, and K. Yabana, 
Phys. Rev. C{\bf 76}, 024318 (2007) ), 
simplifies significantly the fully self-consistent RPA 
calculation. Employing the FAM, we are conducting systematic, 
fully self-consistent response calculations for a wide mass region. 
This paper is intended to present a computational scheme 
to be used in the systematic investigation and to show 
the performance of the FAM for a realistic Skyrme 
energy functional. We implemented the method in the mixed 
representation in which the forward and backward RPA 
amplitudes are represented by indices of single-particle 
orbitals for occupied states and the spatial grid points 
for unoccupied states. We solve the linear response 
equation for a given frequency. The equation is a linear 
algebraic problem with a sparse non-hermitian matrix,
which is solved with an iterative method.
We show results of 
the dipole response for selected spherical and deformed 
nuclei.
The peak energies of the giant dipole 
resonance agree well with measurements for heavy nuclei, 
while they are systematically underestimated for light 
nuclei. We also discuss the width of the giant dipole 
resonance in the fully self-consistent RPA calculation.
\end{abstract}

\maketitle

\section{Introduction}
Nuclear density-functional theory has been successful 
for a systematic description of ground-state properties 
such as binding energies, density distributions, 
deformations, and so on. 
In the last decade, systematic density-functional calculations 
for a whole nuclear chart have come to be an ordinary work 
where a variety of energy functionals are employed \cite{LPT03,Sto03}. 
One of the major goals of such systematic investigations is 
an assessment and an improvement of the energy functionals. 
A rapid progress of experimental research on unstable nuclei 
also necessitates a development of an accurate and universal 
energy functional that describes nuclei far from the stability 
line. These efforts ultimately aim to establish a theory with a
high predictive power and high accuracy.

The nuclear density-functional theory is also capable of 
describing excited state properties. 
The random-phase approximation (RPA), which is derived 
by linearizing the time-dependent mean-field equation 
\cite{RS80}, describes the nuclear excitation as a 
small-amplitude oscillations around the ground state. 
It has been successful for describing both giant 
resonances and low-lying modes of excitations. 
In practical calculations, the RPA equation has been solved 
mostly in the matrix diagonalization scheme.
In these calculations, it was common to ignore a part of
the residual interaction, sacrificing the full self-consistency. 
This is because the inclusion of the full residual interaction 
for a realistic interaction involves a cumbersome and complicated task. 
Recently several groups have reported fully self-consistent 
RPA calculations \cite{Ter05,Ter06,FC05,Sil06,PRNV03}. 
However, they are mostly restricted to spherical nuclei. 
There are a few recent attempts for deformed 
nuclei \cite{PG08,PR08,YG08,IH03,Ina06,NY05}.

Recently, we have proposed a new method to solve the RPA 
equation, the finite amplitude method (FAM) \cite{NIY07}. 
The FAM allows us to evaluate the fully self-consistent 
residual fields as a finite difference, employing a computational 
code for the {\it static} mean-field Hamiltonian alone with 
a minor modification. 
In Ref.\cite{NIY07}, we showed that the FAM works accurately 
in solving the RPA equation for a deformed nucleus $^{20}$Ne 
with a simplified interaction.

Employing the FAM, we are undertaking a systematic investigation 
of nuclear response for a wide mass region ($A \leq 100$). 
The purpose of this paper is to explain the implementation of 
the FAM with a realistic Skyrme functional to be employed 
in such systematic investigation. 

Our scheme is summarized as follows:
\begin{enumerate}
\item Fully self-consistent RPA calculation. The identical 
Skyrme energy functional is used for both ground state and 
linear response calculations, including spin-orbit, Coulomb, 
and time-odd-density terms.
\item Applicable to deformed nuclei. To describe any kind of 
deformation, we employ a three-dimensional (3D) real-space 
grid representation.
\item Linear response equation is solved employing an iterative 
method for the external field with a given frequency. 
\end{enumerate}

We employ a mixed representation in which the RPA forward and 
backward amplitudes are described with orbital indices for 
occupied single-particle states and the 3D grid-points for 
unoccupied states. In the mixed representation, the RPA 
equation is a linear algebraic equation with a sparse, 
non-hermitian matrix. The dimensionality is typically an 
order of $10^6$. We will examine and compare performance of 
different solvers for the linear algebraic problem, and 
demonstrate that the FAM works excellently for realistic 
calculations with complicated energy functionals.

At present, we ignore the pairing terms. The pairing effects 
will not be important for high-lying negative-parity excitations 
such as the dipole excitation which we investigate in this paper. 
However, the pairing may have significant effects on the 
low-lying modes such as quadrupole excitations. An extension 
to include the pairing is now under study. In order to treat the 
particle escape, we need to impose a continuum boundary 
condition \cite{SB75}. Although it can be managed in the present 
scheme \cite{NY05}, the accurate description requires a heavy 
computational task. For a systematic investigation which we 
pursue, we must find a compromise between the accuracy of the 
calculation and the computational feasibility. We will 
approximately take into account the continuum effects by solving 
the RPA equation inside a 3D spherical box of a large radius.

This paper is organized as follows. In Sec. \ref{sec:formalism}, 
we review formalism of the FAM. The numerical details and the 
related techniques are discussed in Sec. \ref{sec:numerical_details}. 
To show the accuracy of the method, the FAM results are compared 
with the fully self-consistent RPA results in the conventional 
diagonalization scheme. Calculated results are shown in 
Sec.~\ref{sec:numerical_results} for selected nuclei in light 
and heavy mass regions. Convergence with respect to the model space, 
dependence on the functionals will be discussed. 
Finally, our fully self-consistent RPA results are compared with 
experimental data. Conclusions are presented in Sec.~\ref{sec:conclusion}.

\section{Linear-response calculation with a Skyrme functional}
\label{sec:formalism}

\subsection{Linear response equation in the mixed representation}

Under a weak, time-dependent external field $V_\mathrm{ext}(t)$, 
the transition density $\delta\rho(t)$ follows the equation 
($\hbar = 1$)
\begin{equation}
\label{TD_linear}
i \frac{\mathrm{d}}{\mathrm{d}t}\delta \rho(t) = \left[ h_0 \,,\, \delta\rho(t)  \right] + \left[ V_\mathrm{ext}(t) + \delta h(t) \,\,  \rho_0  \right] ,
\end{equation}
where $\rho_0$ and $h_0=h[\rho_0]$ are the ground-state density and the 
single-particle Hamiltonian, respectively. 
$\delta h(t)$ is a residual field induced by density fluctuation, 
$h[\rho(t)]=h_0+\delta h(t)$. 
Assuming that $\delta\rho(t)$, $V_\mathrm{ext}(t)$, and $\delta h(t)$ 
oscillate with a frequency $\omega$ 
like $\delta\rho(t)=\delta\rho(\omega)e^{-i\omega t}
+\delta\rho^\dagger(\omega)e^{i\omega t}$, 
Equation (\ref{TD_linear}) is recast to 
\begin{equation}
\omega \delta\rho(\omega) = \left[ h_0 ,\, \delta\rho(\omega) \right]
 + \left[ V_\mathrm{ext}(\omega) + \delta h(\omega) , \rho_0 \right] .
\label{freq.repre.}
\end{equation}
Since $\delta\rho(\omega)$ is not necessarily hermitian, we need 
forward and backward amplitudes, $\ket{X_i(\omega)}$ and 
$\ket{Y_i(\omega)}$, to express the transition density $\delta \rho(\omega)$. 
\begin{equation}
\delta\rho(\omega) = \sum_{i=1}^A
  \left\{ \ket{X_i(\omega)}\bra{\phi_i}
  + \ket{\phi_i}\bra{Y_i(\omega)}  \right\} ,
\label{DeltaRhoOmega}
\end{equation}
where $A=N+Z$ is the mass number of the nucleus 
and the orbitals $\phi_i$ are occupied orbitals in the ground state, 
$h_0 \ket{\phi_i} = \epsilon_i \ket{\phi_i}$ ($i=1,\cdots,A$). 
Substituting this expression into Eq.~(\ref{freq.repre.}), we obtain the 
RPA linear-response equation:
\begin{eqnarray}
\label{RPAeqX}
\omega \ket{X_i(\omega)} &=& \left( h_0 - \epsilon_i \right) \ket{X_i(\omega)}
\nonumber \\ && \hspace{30pt}
 +\hat{P}\left\{ V_\mathrm{ext}(\omega) + \delta h(\omega) \right\} \ket{\phi_i}
, \\
\label{RPAeqY}
-\omega \bra{Y_i(\omega)} &=& \bra{Y_i(\omega)} \left( h_0 - \epsilon_i \right) 
\nonumber \\&& \hspace{30pt}
+ \bra{\phi_i}\left\{ V_\mathrm{ext}(\omega) + \delta h(\omega) \right\}\hat{P}
,
\end{eqnarray}
where $\hat P$ denotes the projector onto the particles space,
$\hat{P} =  1 -  \sum_{i=1}^A | \phi_i \rangle\langle \phi_i |$.
In Eqs.~(\ref{RPAeqX}) and (\ref{RPAeqY}), 
if we expand $\delta h(\omega)$ with respect to
the forward and backward amplitudes in the linear order,
we obtain the well-known $A-B$ matrix form of the RPA equations \cite{NIY07,RS80}.
When the single-particle Hamiltonian and the external field are 
both local in coordinate space, we may write 
Eqs.~(\ref{RPAeqX}) and (\ref{RPAeqY}) conveniently in 
the coordinate representation, 
\begin{eqnarray}
\omega X_i({\xi},\omega) &=& 
\left( h_0({\xi}) - \epsilon_i \right) X_i({\xi},\omega)
\nonumber \\
&+& \hat{P} \left\{ V_\mathrm{ext}({\xi},\omega) 
+ \delta h({\xi},\omega) \right\} \phi_i({\xi}) , \label{RPAeqXr}\\
- \omega Y_i^*({\xi},\omega)  &=&
\left[\left( h_0({\xi}) - \epsilon_i \right) Y_i({\xi},\omega)
\right.
\nonumber \\
&+& \left.
 \hat{P} \left\{ V^\dagger_\mathrm{ext}({\vec \xi},\omega) 
+ \delta h^\dagger({\xi},\omega) \right\}\phi_i({\xi})\right]^* ,
\label{RPAeqYr}
\end{eqnarray}
where the coordinate $\xi$ may contain the spin index $\sigma$ as well 
as the spatial coordinate $\vec{r}$, $\xi=(\vec{r},\sigma)$. 
This is often referred to the mixed representation \cite{LV68,IH03,Ina06}. 
It should be noted that, since $\delta h(\omega)$ has a linear dependence 
on $X_i(\omega)$ and $Y_i^*(\omega)$, 
this is an inhomogeneous linear algebraic equation of the form 
$\mathbf{A}\vec{x}=\vec{b}$ where 
$\vec{x}\equiv (X_i(\xi,\omega),Y_i^*(\xi,\omega))$ \cite{NIY07}. 
In our implementation, we employ the grid representation of the 3D 
Cartesian-coordinate space. 
Denoting the number of the spatial grid points as $N_{\vec{r}}$, 
the dimension of the vector $\vec{x}$ is 
$N_{\vec{r}} \times A \times 2 \times 2$ 
(a factor of two for the forward and backward amplitudes and 
another factor of two for the spin degrees of freedom). 
$N_{\vec{r}}$ is order of 10,000 (see Sec. \ref{sec:adaptive_coordinate}). 
For systems described by local potentials, we need not introduce any  
further truncations in the particle space. In the 3D grid representation 
the treatment of the particle escape is better controlled than that in 
other representations, for example, the harmonic-oscillator-basis 
representation.

\subsection{Finite amplitude method (FAM)}
\label{sec:FAM}

If we write the residual field $\delta h({\xi},\omega)$ 
explicitly with the forward $X_i({\xi},\omega)$ and backward 
$Y_i({\xi},\omega)$ amplitudes, it requires us to calculate 
the residual two-body kernel, 
$v(\xi,\xi')=\delta h(\xi,\omega)/\delta\rho(\xi')$. 
However, if one employs a realistic energy functional, the 
explicit construction of the residual two-body kernel 
involves a complicated coding and a large numerical task. 
For this reason, it has been a common procedure to approximate 
or ignore a part of the residual interaction in most RPA 
calculations. For example, the velocity-dependent 
terms are sometimes replaced with the Landau-Migdal interaction. 
The spin-spin, the spin-orbit, and the Coulomb residual 
interactions are often neglected. 

To achieve a fully self-consistent RPA calculation, 
we believe that the FAM \cite{NIY07} is one of the simplest 
methods, at least, with respect to programming. 
In the FAM, we evaluate $\delta h(\xi,\omega)$ directly from the ground state Hamiltonian $h[\rho]$ 
by the finite difference method, thus avoiding an explicit 
construction of the residual two-body kernel. 
In this section, we recapitulate the essence of the FAM. 
Readers are referred to Ref. \cite{NIY07} for details.

The residual field $\delta h(\xi,\omega)$ depends linearly 
on $\delta\rho(\omega)$:
$
\delta h(\omega) = \partial h/\partial\rho \vert_{\rho=\rho_0}
 \cdot \delta\rho(\omega)
$.
In the first oder with respect to a small parameter $\eta$, we have 
\begin{equation}
h_0 + \eta \delta h(\omega) =
h\left[ \rho_0 + \eta \delta\rho(\omega) \right] \equiv
h\left[ \tilde\rho_\eta(\omega) \right] .
\label{linearH}
\end{equation}
Here, the one-body {\it pseudo-density} operator 
$\tilde\rho_\eta(\omega)\equiv \rho_0+\eta\delta\rho(\omega)$ 
is a non-hermitian operator expressed with bra and ket vectors 
as follows:
\begin{equation}
\tilde\rho_\eta(\omega)= \sum_i \left( | \phi_i \rangle + \eta | X_i(\omega) \rangle \right) \left( \langle \phi_i | + \eta \langle Y_i(\omega) | \right).
\end{equation}
In the linear order in $\eta$, Equation (\ref{linearH}) is expressed as
\begin{equation}
\delta h(\omega) = \frac{1}{\eta} \left( h\left[ \tilde\rho_\eta(\omega)\right]
                                       - h_0 \right) .
\label{FAM}
\end{equation}
This is the basic result of the FAM. 
Equation (\ref{FAM}) indicates that 
$h\left[ \tilde\rho_\eta(\omega)\right]$ can be evaluated in the 
same manner as in $h_0=h[\rho_0]$, with a replacement of the 
ordinary densities with the {\it pseudo-densities}:
For instance, the {\it pseudo-local-density}, {\it pseudo-kinetic-density}, and 
{\it pseudo-current-density } are given by 
\begin{eqnarray}
&&\tilde\rho_\eta(\vec{r})=\sum_i \sum_\sigma \psi_i^*(\xi) \varphi_i(\xi), \nonumber\\
&&\tilde\tau_\eta(\vec{r})=\sum_i \sum_\sigma \nabla \psi_i^*(\xi) \cdot
                            \nabla \varphi_i(\xi), \nonumber \\
&&\vec{\tilde j}_\eta(\vec{r})=\frac{1}{2i}
     \sum_i \sum_\sigma \{ \psi_i^*(\xi) \nabla \varphi_i(\xi)
                - (\nabla \psi_i^*(\vec{r})) \varphi_i(\vec{r}) \},
		\nonumber\\
\label{densities}
\end{eqnarray}
respectively, 
where $\varphi_i=\phi_i + \eta X_i(\omega)$ and 
$\psi_i=\phi_i + \eta Y_i(\omega)$. 
The spin-dependent pseudo-densities are also defined 
in the same manner. Once $X_i(\omega)$, $Y_i(\omega)$, 
and $\eta$ are given, the pseudo-densities are calculated 
as in Eq.~(\ref{densities}). 

Equations (\ref{RPAeqXr}) and (\ref{RPAeqYr}) are the linear 
algebraic equation of the form, 
$\mathbf{A} \vec{x} = \vec{b}$. 
Here,
\begin{equation}
\vec{x}=
\begin{pmatrix}
X_i(\xi,\omega) \\
Y_i^*(\xi,\omega)
\end{pmatrix}
,
\quad\quad
\vec{b}=
\begin{pmatrix}
-V_\mathrm{ext}(\xi,\omega) \phi_i(\xi)\\
-\left\{ V_\mathrm{ext}(\xi,\omega) \phi_i(\xi) \right\}^*
\end{pmatrix}
.
\end{equation}
To solve Eqs. (\ref{RPAeqXr}) and (\ref{RPAeqYr}), 
we utilize the iterative scheme, as described below. 
In the iterative scheme, we need not construct the matrix elements 
of $\mathbf{A}$ explicitly but always evaluate $\mathbf{A}\vec{x}$ 
employing the FAM for a given vector $\vec{x}$. 
\begin{equation}
\mathbf{A}\vec{x}=
\begin{pmatrix}
(h_0(\xi)-\epsilon_i-\omega) X_i(\xi,\omega)
  +\delta h(\xi,\omega) \phi_i(\xi) \\
\left\{ (h_0(\xi)-\epsilon_i+\omega^*) Y_i(\xi,\omega)
  +\delta h^\dagger(\xi,\omega) \phi_i(\xi) \right\}^*\\
\end{pmatrix}
,
\end{equation}
where $\delta h(\xi,\omega) \phi_i(\xi)$ is calculated 
using Eq. (\ref{FAM}) and 
$\delta h^\dagger(\xi,\omega) \phi_i(\xi)$ 
using the same equation but 
$\tilde\rho_\eta$ replaced by $\tilde\rho_\eta^\dagger$. 
To find a solution of Eqs. (\ref{RPAeqXr}) and (\ref{RPAeqYr}), 
we employ extensions of the conjugate gradient method for 
linear algebraic equations involving a non-hermitian matrix. 
In the literature, quite a few solvers for linear algebraic 
equations involving a non-hermitian matrix have been developed. 
However, we find only a few methods work for the present problem. 
We will discuss it in Sec.~\ref{sec:numerical_details}.

In order to calculate the strength function for a given 
one-body operator $F$, we adopt an external field of 
$V_\mathrm{ext}(t) = \eta F e^{-i \omega t}+ \eta^\ast F^\dag e^{i \omega t}$. 
Then, the transition strength is expressed with the 
forward and backward amplitudes, 
\begin{eqnarray}
\frac{dB(E;F)}{dE} 
&\equiv&  -\frac{1}{\pi} \mathrm{Im} 
  \sum_i \left\{ \bra{\phi_i} F^\dagger \ket{X_i(\omega)} +
           \bra{Y_i(\omega)} F^\dagger \ket{\phi_i} \right\} \nonumber \\
&=& \sum_n \left|\bra{n}F\ket{0} \right|^2 \delta(E-E_n),
\label{dB/dw}
\end{eqnarray}
for a real frequency $\omega=E$. 
Here, $\ket{n}$ are energy eigenstates of the total system. 
In the following calculations, we use 
complex frequencies with a finite imaginary part, 
$\omega =E + i \gamma /2$.
Then, the transition strength becomes 
\begin{eqnarray}
\frac{dB(E;F)}{dE} 
= \frac{\gamma}{2\pi} \sum_n \left\{
\frac{ | \bra{n} F \ket{0} |^2 }
     {\left( E - E_n \right)^2 + \left( \gamma /2 \right)^2}
\right.
\nonumber \\
\left.
- \frac{ | \bra{n} F^\dag \ket{0}|^2 }
       {\left( E + E_n \right)^2 + \left( \gamma /2 \right)^2}
 \right\} 
\label{FAM_smearing}
\end{eqnarray}
The second term in the right hand side is not important, if 
$\gamma \ll E+E_n$. 
For an hermitian operator $F$, this leads to 
\begin{equation}
\frac{dB(E;F)}{dE} 
= \frac{2E\gamma}{\pi} \sum_n
\frac{ \tilde{E}_n | \bra{n} F \ket{0} |^2 }
     {\left( E^2 - \tilde{E}_n^2 \right)^2 + E^2\gamma^2 } ,
\label{dB/dw_gamma}
\end{equation}
where $\tilde{E}_n^2\equiv E_n^2 + \gamma^2/4$.

In this article, we consider an electric dipole operator 
for $F$. 
\begin{equation}
D^{E1}_z =\frac{N}{A}e \sum^Z_{p=1} z_p
      -\frac{Z}{A}e \sum^N_{n=1} z_n ,
\label{D_E1}
\end{equation}
and similar operators for $D^{E1}_x$ and $D^{E1}_y$. 
The photoabsorption cross section in the dipole 
approximation is given as follows \cite{BM75}, 
\begin{equation}
\sigma_\mathrm{abs}(E)= \frac{4\pi^2E}{3 c}
\sum_{\mu=x,y,z} \frac{dB(E; D^{E1}_\mu)}{dE} .
\label{sigma}
\end{equation}

\subsection{Numerical details}
\label{sec:numerical_details}

\subsubsection{Adaptive 3D grid representation}
\label{sec:adaptive_coordinate}

We employ a model space of 3D grid points inside 
a sphere of a radius $R_\mathrm{box}$. 
All the single-particle wave functions and potentials 
except for the Coulomb potential are assumed to 
vanish outside the sphere. For the calculation of 
the Coulomb potential, we follow the prescription 
in Ref. \cite{FKW78}. 
The differentiation is approximated by a finite 
difference with the nine-point formula.

In order to obtain nuclear response with a reasonable 
accuracy, we need a large 
model space, typically $R_\mathrm{box}\gtrsim 15$ fm
(see Sec.~\ref{sec:box_size}).
This is much larger than the typical nuclear size $R_0$.
Outside the nuclear radius $R_0$, we do not need to use 
a fine mesh. Thus we adopt an adaptive coordinate system 
to reduce the number of grid points in the outer
region, $R_0 < r < R_\mathrm{box}$. We use the 
following coordinate transformation, $(u,v,w) \to (x,y,z)$, 
which was also adopted in Ref.~\cite{NY05},
\begin{equation}
\label{x_u}
x(u) = ku \left[ 1 + ( k-1 )
      \left\{ \frac{u}{x_0\sinh ( u/ x_0 )} \right\}^n\right]^{-1} ,
\end{equation}
and the same form for $y(v)$ and $z(w)$.
In this transformation, we have $x(u)\approx u$ for spatial region
of $u \ll x_0$, and $x(u)\approx ku$ for $u \gg x_0$.
We adopt a uniform mesh spacing of $\Delta h=0.8$ fm
in the $(u,v,w)$ space.
The parameters, $k=5.0$, $x_0=8$ fm, and $n=2$, are used in the
following calculations.
The number of grid points in the sphere of $r \le R_\mathrm{box}$ is
significantly reduced; $27,609 \to 11,777$ for $R_\mathrm{box}=15$ fm
and $221,119 \to 39,321$ for $R_\mathrm{box}=30$ fm.

\subsubsection{Choice of Iterative algorithms}
\label{sec:algorithm}

In this section, we discuss performance of different iterative 
solvers to solve the linear response equations. 
We first discuss the calculation for the ground state. 
To obtain ground-state solutions,
the imaginary-time iterative method is used \cite{DFKW80}.
We impose constraints in the 
iteration process so that the center-of-mass always coincides 
with the origin of the coordinate system and the principal axes 
of the density distribution coincide with three Cartesian axes 
of $x$, $y$, and $z$.
It is important to obtain a well converged ground state
solution, since the convergence properties of the 
linear-response calculation crucially depends on this.
To assure the strict convergence in the ground state
calculation, we impose the convergence condition not only for 
the total energy, but also for the deformation 
parameters $(\beta,\gamma)$, single-particle 
energies $\epsilon_i$, single-particle angular momenta, 
and so on.

\begin{figure}[tb]
\begin{center}
\includegraphics[width=0.45\textwidth,keepaspectratio]{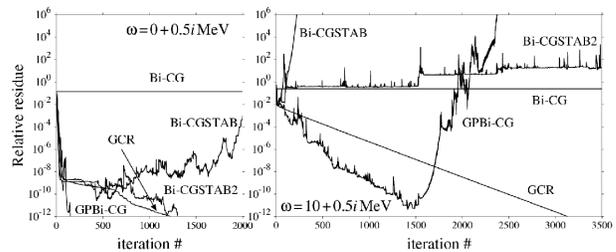}
\caption{{\small 
Convergence property of different iterative methods to solve the linear 
response equations~(\ref{RPAeqXr}) and (\ref{RPAeqYr}) for
electric dipole response in $^{16}$O calculated at complex frequencies of
$\omega=0+0.5i$ MeV (left) and $10+0.5i$ MeV (right).
Relative residue, $r_n$,
is shown as a function of iteration number $n$.
}}
\label{fig:convergence_property}
\end{center}
\end{figure}

For the Skyrme energy functional, the RPA matrix $\mathbf{A}$ which 
appears in Eqs.~(\ref{RPAeqXr}) and (\ref{RPAeqYr}) 
is sparse in the $\vec{r}$-space grid representation.
Therefore, the iterative methods, such as the conjugate gradient 
(CG) method \cite{PTVF07}, should work efficiently.
The CG method is very powerful for the hermitian matrix.
However, since we calculate for the complex frequency $\omega$, 
the matrix $\mathbf{A}$ is no longer hermitian. Therefore, a simple CG 
method is not applicable.
There are a lot of variants of the CG method extended for non-hermitian
problems. We test some of them: Bi-conjugate gradient (Bi-CG) method 
\cite{PTVF07}, generalized conjugate residual (GCR) method 
\cite{EES83}, generalized product-type bi-conjugate gradient 
(GPBi-CG) method \cite{Zhang97}, Bi-CGSTAB method \cite{Vorst92}, 
and Bi-CGSTAB2 method \cite{Gutk93}.
In the original GCR method, it is necessary to store all the vectors
$\vec{x}_0,\cdots, \vec{x}_{n-1}$ in order to calculate the vector 
at the $n$-th iteration. However, this is impractical because it
requires heavy computational task and huge memory size. To avoid the
problem, we restart the GCR procedure every twenty iterations.
In Fig.~\ref{fig:convergence_property}, we show the convergence 
behavior of the different iterative solvers.
The magnitude of the relative residue,
\begin{equation}
r_n = \vert \vec{b}-\mathbf{A}\vec{x_n} \vert 
/ \vert \vec{b} \vert
\label{residue}
\end{equation}
is plotted against the number of iterations.

At very low energies ($\omega=0+0.5i$ MeV),
all the solvers except for the Bi-CG method provide quickly converged
results. On the other hand, at higher energies ($\omega=10+0.5i$ MeV), 
only the GCR and the GPBi-CG methods reach the convergence.
In most cases, the convergence of the GPBi-CG method is faster than
the GCR. However, after the convergence, the residue by the GPBi-CG 
method start to increase again. Therefore, only the result of the GCR 
method shows a stable convergence property 
in Fig.~\ref{fig:convergence_property}.
We thus employ the GCR in the following calculations, although it 
has a disadvantage that it requires larger computer memory size
than other methods based on the Bi-CG.

In the iteration procedure, we need to set up the initial vector
$\vec{x}_0$. It turns out that the convergence property depends very little 
on the selection of the initial vector.
We simply take the initial vector of
$\vec{x}_0=0$.
We stop the GCR iteration procedure if the relative residue becomes 
smaller than a threshold, $r_n < 10^{-6}$.
We find the typical number of the iteration necessary to reach 
the convergence is about $10^3$ around the excitation energy region 
of giant dipole resonance (GDR), while the convergence is much faster
for smaller $E(=\mathrm{Re}\,\omega)$. The iteration number to reach
the convergence also depends on the imaginary part of the frequency,
$\gamma(=2\mathrm{Im}\,\omega)$, and is roughly proportional to
$1/\gamma$.

We will later show how the convergence depends on other
parameters. In the inset of Fig. \ref{Rdep},
we show the residue, $r_n$ of Eq. (\ref{residue}),
as a function of the CPU time
for calculations employing different spatial size $R_{\rm box}$.
We can see that the computational time scales linearly both
with the number of grid points and with the particle number $A$.

In Sec.~\ref{sec:numerical_results}, the calculation is performed
with the following settings, unless otherwise specified:
The energy range
of $0\leq E\leq 38.1$ MeV, discretized in 
spacing of $\Delta E=0.3$ MeV (128 points).
The imaginary part of the frequency is fixed at 0.5 MeV,
corresponding to $\gamma=1.0$ MeV.
The calculated strength is interpolated using the cubic 
spline function.
In most cases, we have utilized PC cluster systems with 
either 64 or 128 processors in parallel.

\subsubsection{Choice of the FAM parameter $\eta$}

The FAM parameter $\eta$ in Eq.~(\ref{FAM}) should be as small as
possible to validate the linearity.
In practice, there is a lower limit to avoid the round-off error.
In the numerical computation,
we use real variables with double precision (8 bytes) and
complex variables with $8\times 2$ bytes.
We use the following value for $\eta$ which differ for every 
iteration to ensure the linearity \cite{NIY07}:
\begin{equation}
\eta = \frac{\eta_\mathrm{Lin}}{\max\{N(X),N(Y)\}} , \quad
N(\psi) = \frac{1}{A}\sqrt{ \sum_{i=1}^A \langle \psi_i | \psi_i \rangle}
\end{equation}
with $\eta_\mathrm{Lin}= 10^{-4}$.
The convergence property of the GCR iteration is not sensitive 
to the value of $\eta_\mathrm{Lin}$, as far as the value is 
in the range of $\eta_\mathrm{Lin} = 10^{-2}-10^{-6}$. 
However, if we use too small value, $r_n$ oscillates in the iteration 
and never reaches the convergence.

\section{Calculated results}
\label{sec:numerical_results}

\begin{figure}[tb]
\begin{center}
\includegraphics[width=0.40\textwidth,keepaspectratio]{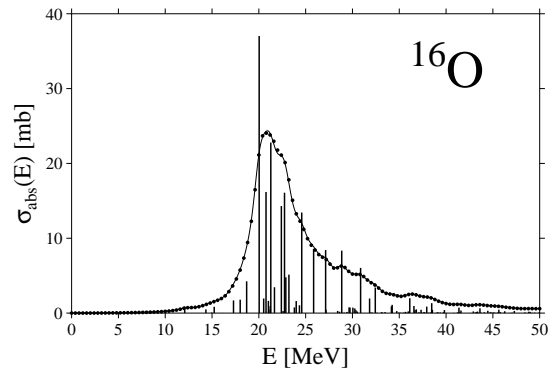}
\caption{{\small 
Comparison of the results for electric dipole response of 
$^{16}$O calculated by the present linear response formalism
with FAM (dots) and by the eigenvalue formalism 
with an explicit construction of the residual interaction (curve).
Both calculations use
the cubic box of (21 fm)$^3$ and
the SIII functional.
See text for details.
The vertical lines indicate eigenenergies and strengths of 
the RPA normal modes calculated by the matrix diagonalization
method,
whose vertical magnitude is in units of mb$\cdot$MeV.
}}
\label{FAMvsRPA}
\end{center}
\end{figure}

\subsection{Numerical accuracy}

In this section, we demonstrate that the FAM, described in
Sec.~\ref{sec:formalism},
really works as a simple and accurate method for
the fully self-consistent linear-response calculation
with realistic Skyrme functionals. 
In Ref. \cite{NIY07}, we have already done it for a simple 
Bonche-Koonin-Negele interaction \cite{BKN76}.

In the following calculations, we use the Skyrme energy functional.
We adopt the same functional form as that of Ref.~\cite{BFH87}
(Appendix A).
In Ref.~\cite{BFH87}, every single-particle orbital is assumed 
to have a definite parity and $z$-signature symmetries. However, 
we do not assume any of these in this paper.
The functional depends on densities $\rho_\tau(\vec{r})$,
kinetic densities $\tau_\tau(\vec{r})$,
spin-orbit densities $\vec{J}_\tau(\vec{r})$,
current densities $\vec{j}_\tau(\vec{r})$,
and spin densities $\vec{\rho}_\tau(\vec{r})$,
where $\tau=n$ and $p$.
The exchange part of the Coulomb energy is evaluated by
the Slater approximation.
The center-of-mass correction is achieved by using
the recoil nucleon mass, $M\to AM/(A-1)$.

In Fig. \ref{FAMvsRPA}, we show photoabsorption 
cross section for $^{16}$O calculated by two different 
methods: The dots are obtained by solving the
linear response equations (\ref{RPAeqXr}) and (\ref{RPAeqYr})
in which the FAM is employed to evaluate the residual
field. The vertical lines are obtained by the matrix
diagonalization method \cite{IH03, Ina06} in which the 
residual interaction taking a second derivative of the 
Skyrme energy function is constructed explicitly.
The positions of the vertical lines indicate the energy
eigenvalues, while the heights show the calculated
cross section. The solid curve is obtained by smoothing
the vertical lines with the energy-weighted Lorentzian.
In both calculations of the linear response and the
matrix diagonalization, the same Skyrme functional and
the same grid representation of the $\vec{r}$-space
are used. The width parameter is set common, $\gamma=2$ MeV.
In the matrix diagonalization method, the spurious 
solutions corresponding to the center-of-mass motion
appear around a few tens keV. They are not shown in the 
figure.

This figure clearly demonstrates that the linear response
calculation with the FAM denoted by dots completely 
agrees with the fully self-consistent RPA calculation with 
an explicit construction of the residual interaction.
We should note that the computer program of the 
diagonalization method requires heavy and complicated 
coding for the residual interaction, while the program coding
with the FAM can be achieved with the static mean-field
calculation alone, with a minor modification. 
Thus the accuracy and the simplicity of the FAM is clearly
demonstrated.

Let us next consider the merit of the linear response 
calculation for a fixed frequency in comparison with the matrix 
diagonalization method, apart from the FAM. 
In the matrix diagonalization method,
the number of normal modes increases rapidly as the
excitation energy becomes higher. Even when we assume the parity 
and the $z$-signature symmetries, the number of excited states 
is an order of one thousand in the energy range below 50 MeV.
If we achieve matrix diagonalization in the mixed representation,
we must calculate eigenvalues and eigenvectors of the
RPA solution one by one. Then the calculations of a thousand
of eigenstates require huge computational time.
Therefore, to calculate responses for a wide energy region,
the computation time with the linear response method is 
much smaller than that of the diagonalization method.
Furthermore, we can easily parallelize the calculation in
the linear response calculation, assigning different 
processors to solve the equation at different frequencies $\omega$.

Combining these features, we consider the linear response 
calculation with the FAM is the best choice for
the systematic, fully self-consistent calculations over a periodic 
table.

\subsection{Light nuclei}

\subsubsection{Dependence on box size and smoothing parameter}
\label{sec:box_size}

\begin{figure}[!tb]
\begin{center}
\includegraphics[width=0.48\textwidth,keepaspectratio]{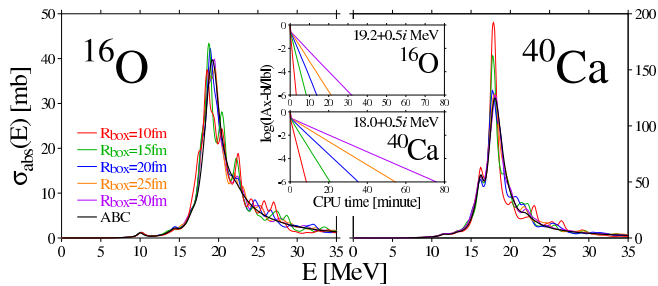}
\caption{{\small 
Photoabsorption cross sections in $^{16}$O and $^{40}$Ca 
calculated with the FAM in a spherical box of different sizes:
$R_\mathrm{box}=10$ fm, 15 fm, 20 fm, and 30 fm.
``ABC'' indicates a result with the absorbing boundary condition 
which properly treats the continuum effect \cite{NY05}.
The SkM$^\ast$ 
interaction and the smoothing parameter $\gamma=1$ MeV are used.
Inset: Convergence properties as a function of the CPU time.
}}
\label{Rdep}
\end{center}
\end{figure}

\begin{figure}[!tb]
\begin{center}
\includegraphics[width=0.48\textwidth,keepaspectratio]{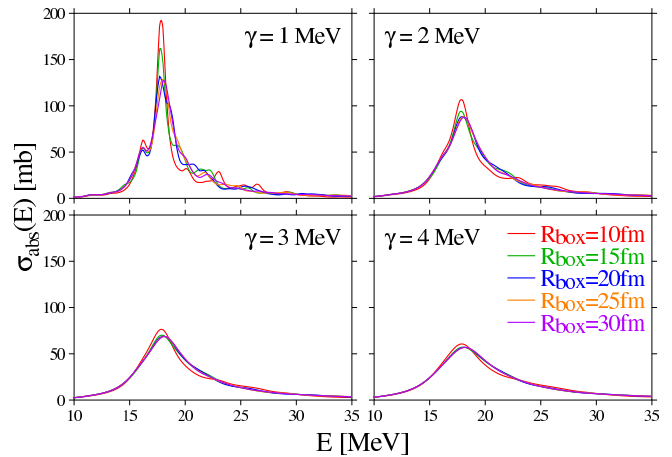}
\caption{{\small 
Photoabsorption cross sections in spherical nucleus $^{40}$Ca 
calculated with different $R_{\rm box}$ and with different $\gamma$ values.
The functional parameter set of SkM$^*$ is used.
}}
\label{Rdep.gammadep.40Ca}
\end{center}
\end{figure}

\begin{figure}[!tb]
\begin{center}
\includegraphics[width=0.48\textwidth,keepaspectratio]{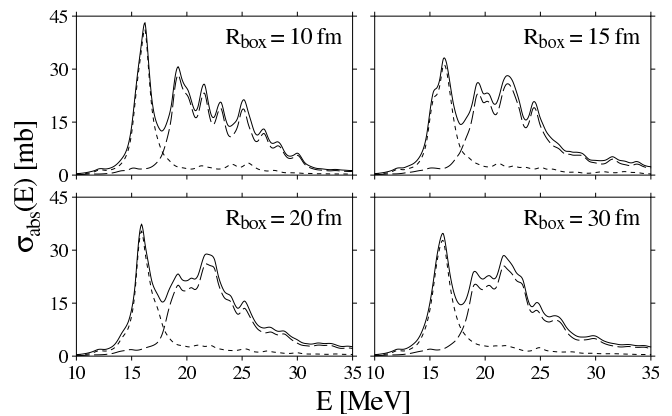}
\caption{{\small 
Photoabsorption cross sections in deformed nucleus $^{24}$Mg 
calculated with a spherical box of 
$R_\mathrm{box}=10$ fm, 15 fm, 20 fm, and 30 fm. Dashed, long dashed, 
and solid lines correspond to cross sections of $K=0$ mode, 
$K=1$ mode, and the total one, respectively.
The functional parameter set of SkM$^*$ and the smoothing parameter
$\gamma=1$ MeV are used.
}}
\label{Rdep.24Mg}
\end{center}
\end{figure}

We show how the results depend on
the box size $R_\mathrm{box}$ and the smoothing parameter $\gamma$.
In the present mixed representation, the model space is specified
with $R_\mathrm{box}$ and the 3D mesh size $\Delta h$.
The dependence on the mesh size is discussed in details in
Ref.~\cite{IH03} for spurious and low-energy excitations
in $^{16}$O and $^{208}$Pb.
According to this study, $\Delta h\approx 0.8$ fm produces a
converged result.
In addition, we have also carefully examined the convergence 
with respect to the grid spacing in the adaptive-coordinate.
We adopt $x_0=8$ fm in Eq.~(\ref{x_u}).

In Fig. \ref{Rdep}, results of the calculations with
$R_\mathrm{box}=10$, 15, 20, 25, and 30 fm are shown.
As is seen in Fig. \ref{Rdep},
small peaks in higher energies ($E > 20$ MeV) are disappearing
with increasing $R_\mathrm{box}$.
These small peaks at high energies are all spurious,
due to the discretization of the continuum.
In fact, the calculation with the proper continuum boundary
condition \cite{NY05}, which is denoted as ``ABC'', shows only
a smooth tail in this energy region above the main GDR peak.
With the energy resolution of $\gamma=1$ MeV,
the effect of the discretized continuum is still visible at $E>20$ MeV,
even for the case of $R_\mathrm{box} \simeq 30$ fm.
In order to remove the spurious effect, we need increase either
$R_\mathrm{box}$ or $\gamma$.
In Fig. \ref{Rdep.gammadep.40Ca}, we show results for $^{40}$Ca calculated
with different box sizes and with different magnitudes of the
smoothing parameter $\gamma$.
Using $\gamma>2$ MeV,
the calculation roughly converges with $R_{\rm box}>15$ fm.
Thus, in the calculation with the
vanishing (box) boundary condition, it is difficult to distinguish
small physical peaks from unphysical ones in the high-energy
continuum region, $E > 20$ MeV.
In this article, we concentrate our discussion on gross structure of
the main GDR peaks.

Next, in Fig. \ref{Rdep.24Mg}, we show results for
$^{24}$Mg. Since the ground state of $^{24}$Mg is deformed with a prolate
shape ($\beta_2= 0.39$), the GDR splits into two major peaks, 
$K=0$ mode (dashed line) and $K=1$ mode (long dashed line).
The $K=0$ mode corresponds to the oscillation parallel to the symmetry
axis ($z$ direction) and $K=1$ mode to that perpendicular to the symmetry axis
($x$ and $y$ directions).
In the calculations with $R_\mathrm{box}=10$ fm and 15 fm,
we observe a peak at $E=25$ MeV for the $K=1$ mode.
However, this peak becomes smaller for larger $R_\mathrm{box}$.
This indicates that the peak at 25 MeV comes from the effect of the
box discretization and is spurious.

\begin{table}
\caption{
$R_{\rm box}$-dependence of the GDR peak positions and widths in units of MeV,
for $^{16}$O and $^{40}$Ca.
The smoothing parameter of $\gamma=1$ MeV is used.
See text for details.
}
\label{table.Rdep}
\begin{center}
\begin{tabular}{|c|c|c||c|c|}
\hline
 & \multicolumn{2}{|c||}{$dB(E)/dE$} 
 & \multicolumn{2}{|c|}{$\sigma_\mathrm{abs}(E)$} \\ \cline{2-5}
$^{16}$O  & $E_\mathrm{peak}$ & $\Gamma$ 
          & $E_\mathrm{peak}$ & $\Gamma$  \\
\hline
 $R_\mathrm{box} = 10$ fm  &  18.993 &  3.869  &  18.798 &  3.876 \\
 $R_\mathrm{box} = 15$ fm  &  19.063 &  3.475  &  19.103 &  3.944 \\
 $R_\mathrm{box} = 20$ fm  &  19.141 &  3.096  &  19.151 &  3.435 \\
 $R_\mathrm{box} = 30$ fm  &  19.162 &  3.035  &  19.165 &  3.330 \\ \hline
 & \multicolumn{2}{|c||}{$dB(E)/dE$} 
 & \multicolumn{2}{|c|}{$\sigma_\mathrm{abs}(E)$} \\ \cline{2-5}
$^{40}$Ca & $E_\mathrm{peak}$ & $\Gamma$ 
          & $E_\mathrm{peak}$ & $\Gamma$  \\
\hline
 $R_\mathrm{box} = 10$ fm &  17.794 &  1.545 &  17.766 &  1.550 \\
 $R_\mathrm{box} = 15$ fm &  17.797 &  2.334 &  17.796 &  2.492 \\
 $R_\mathrm{box} = 20$ fm &  17.904 &  2.789 &  17.901 &  2.886 \\
 $R_\mathrm{box} = 30$ fm &  17.963 &  2.956 &  17.959 &  3.014 \\ \hline
\end{tabular}
\end{center}
\end{table}

\begin{figure}[tb]
\begin{center}
\includegraphics[width=0.4\textwidth,keepaspectratio]{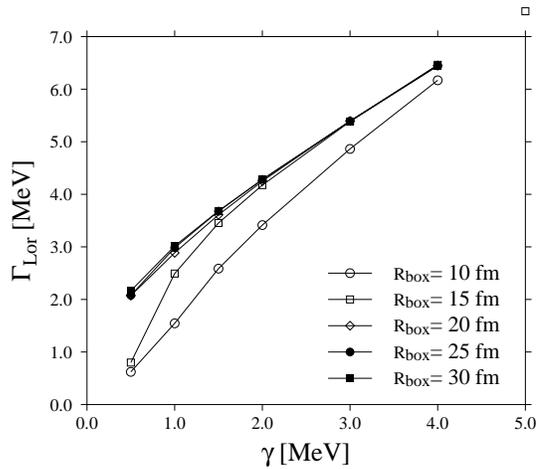}
\caption{{\small 
Estimated GDR width for $^{40}$Ca as a function of the parameter $\gamma$.
}}
\label{gamma-Gamma}
\end{center}
\end{figure}

Although the calculated profile functions of cross section
in the high-energy continuum region 
are affected by the box discretization,
the gross property of the GDR is less sensitive to the value of
$R_\mathrm{box}$.
We estimate the GDR peak energy and the width according to
the electric dipole ($E1$) strength distribution
and the photoabsorption cross section (oscillator strength distribution).
For spherical nuclei, the nuclear response does not depend on
the direction of the external dipole field.
Thus, we can arbitrary choose it, for instance, $F=D_z^{E1}$.
Then, $E1$ strength is fitted by a single Lorentzian function,
\begin{eqnarray}
\frac{dB(E; D_z^{E1})}{dE}\approx
B_\mathrm{tot}\times L(E), \nonumber \\
L(E)=\frac{1}{\pi}
\frac{\Gamma/2}{\left( E - E_\mathrm{peak} \right)^2 + ( \Gamma/2 )^2} ,
\label{standard_Lorentzian}
\end{eqnarray}
where $E_\mathrm{peak}$ and $\Gamma$ are fitting parameters, and
$B_\mathrm{tot}=\int dE dB(E)/dE$.
The fitting is also performed for the photoabsorption cross section
$\sigma(E)$,
using the energy-weighted Lorentzian, $f_\mathrm{tot}\times E L(E)$.
The results are shown in Table \ref{table.Rdep}.
These two definitions of the peak energy and the width lead to
nearly identical results.

The dependence of
$E_\mathrm{peak}$ on the box size $R_\mathrm{box}$ is very small.
On the other hand, the GDR width $\Gamma$ shows a substantial 
dependence on the box size.
For $\gamma=1$ MeV, to estimate the width with a reasonable accuracy
requires $R_\mathrm{box}>20$ fm.
However, the width also depends on the smoothing parameter $\gamma$,
which is shown in Fig.~\ref{gamma-Gamma}.
The value of $\Gamma$ is converged at $R_{\rm box}\geq 20$ fm for
$\gamma=1$ MeV.
When we use a larger $\gamma$ value, the model space of $R_{\rm box}=15$ fm
is good enough to calculate the width.
From the approximate linear dependence of $\Gamma$ on the parameter $\gamma$, 
we can estimate the width at $\gamma=0$ as $\Gamma^{\rm RPA}\approx 2$ MeV.
This value can be regarded as the damping width in the RPA, 
namely due to the Landau damping and escaping:
$\Gamma^{\rm RPA}=\Gamma^{\rm Landau}+\Gamma^\uparrow$.
Since the smoothing parameter $\gamma$ can be regarded as the spreading width,
$\gamma=\Gamma^\downarrow$,
we may extract $\Gamma^\downarrow$ from experimental data.
This will be discussed in the followings.

\subsubsection{Comparison with experiments and Skyrme-functional dependence}

\begin{figure}[tb]
\begin{center}
\includegraphics[width=0.48\textwidth,keepaspectratio]{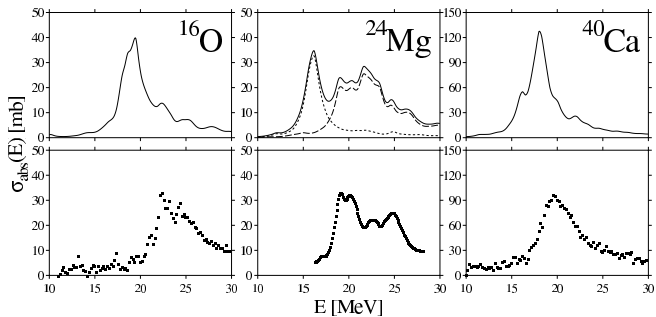}
\caption{{\small 
Calculated (upper panels) and experimental (lower panels)
\cite{Ishkhanov02,Varlamov03} photoabsorption cross sections in spherical 
nuclei $^{16}$O, $^{40}$Ca and deformed nucleus $^{24}$Mg.
We use the box size $R_\mathrm{box}=30$ fm and
the SkM$^\ast$ parameter set with $\gamma=1$ MeV.
}}
\label{fig.comp.exp.light}
\end{center}
\end{figure}

Calculated photoabsorption cross sections for spherical nuclei,
$^{16}$O, $^{40}$Ca and deformed nucleus $^{24}$Mg,
are compared with experimental data
in Fig. \ref{fig.comp.exp.light}. 
The calculated energies of the GDR peaks are underestimated
by a few MeV, but the overall 
structures are reproduced.
For spherical nuclei, the 
GDR widths calculated with $\gamma=1$ MeV are narrower than
the corresponding experimental data.
This seems to suggest that
the spreading width $\Gamma^\downarrow$,
which takes account of effects decaying into compound states,
such as two-particle-two-hole excitations,
is slightly larger than $\gamma=1$ MeV.
The cross section at higher energies is larger in experiment.
This may be due to effect of the ground-state correlation
produced by the tensor and short-range parts in the two-body
nuclear interaction \cite{BM75}.
For $^{16}$O, all the calculations fail to reproduce
a double-peak structure of the GDR, present in experimental data.
If we neglect the time-odd spin densities, the double peaks can be
reproduced \cite{NY05}.
The GDR peak energy in $^{40}$Ca is better reproduced compared to the case
for $^{16}$O.
The calculated 
cross sections in the high-energy region also becomes closer to
the experimental data.
It seems that the discrepancy is more prominent for lighter nuclei.

For the deformed nucleus $^{24}$Mg, the GDR 
peak splitting caused by the ground-state deformation well agree with
the experiments, although the magnitude of the deformation splitting is
slightly too large in the calculation.
We may interpret that the experimental GDR 
peak around $E=20$ MeV is associated with the $K=0$ mode,
and those at $E=22\sim 25$ MeV correspond to the $K=1$ mode.
The double-peak structure of the $K=1$ GDR peak is well reproduced.
Approximately, the calculated cross section is shifted to lower energy
from the experimental ones, by about 3 MeV.

\begin{figure}[tb]
\begin{center}
\includegraphics[width=0.48\textwidth,keepaspectratio]{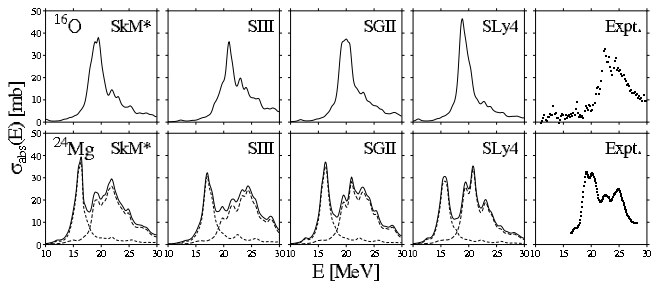}
\caption{{\small 
Comparison of the calculated photoabsorption cross sections with different
Skyrme parameter sets and experimental data in $^{16}$O and $^{24}$Mg.
The calculation is performed with the box size $R_\mathrm{box}=25$ fm. 
}}
\label{fig.comp.exp.24Mg}
\end{center}
\end{figure}

We next show results obtained with different
Skyrme-parameter sets in Fig. \ref{fig.comp.exp.24Mg};
SkM$^\ast$ \cite{Bart82}, SIII \cite{BFGQ75},
SGII \cite{GS81}, and SLy4 \cite{Chan97}.
Two parameter sets, SkM$^\ast$ and SLy4, produce similar results.
In $^{16}$O, the GDR peak position is calculated around $E=19$ MeV.
In contrast, the SIII functional produces the GDR peak near
$E=21$ MeV. The result with the SGII functional is intermediate,
around $E=20$ MeV. However, all the calculations underestimate the experimental
GDR peak energy, $E\approx 22$ MeV.
This discrepancy on the GDR peak energy will disappear for heavier nuclei
(see Sec. \ref{sec:heavy_nuclei}).
The cross section at higher energies $E>25$ MeV is also
systematically underestimated in the calculations.
For deformed $^{24}$Mg,
all the calculations again underestimate the peak energies.
Both of $K=0$ and $K=1$ peaks obtained with 
the SIII functional are located highest in energy compared to the others.

\subsubsection{Transition density}
\begin{figure}[tb]
\begin{center}
\includegraphics[width=0.48\textwidth,keepaspectratio]{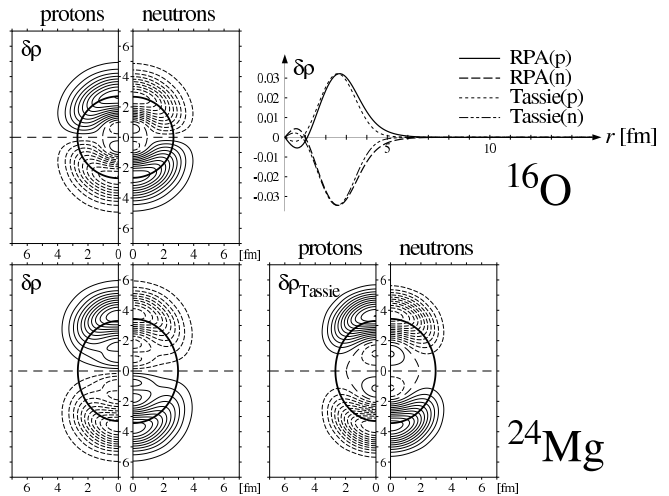}
\caption{{\small 
Left panels: Contour maps of the calculated transition densities of $K=0$ mode,
$\mathrm{Im}\, \delta\rho(\omega)$, in the $y$-$z$ ($x=0$) plane
at $\omega=19.4 + 0.5 i$ MeV in $^{16}$O (upper panels) and $16.1 + 0.5 i$ MeV
in $^{24}$Mg (lower).
Contour lines with positive, negative, and zero values are 
denoted by solid, dashed, and long-dashed lines, respectively.
Thick solid lines correspond to nuclear surface, defined by
$y^2/\langle y^2 \rangle + z^2/\langle z^2 \rangle=1$.
Right panels: Transition densities predicted by the Tassie model.
The magnitude is fitted to the RPA result at the maximum value.
See text for details.
}}
\label{drho}
\end{center}
\end{figure}

We obtain the forward and backward amplitudes, $X_i(\xi,\omega)$
and $Y_i(\xi,\omega)$, using the FAM.
Then, the local part of the transition density in the intrinsic frame,
Eq.~(\ref{DeltaRhoOmega}),
can be calculated as
\begin{equation}
\delta\rho_\tau(\vec{r},\omega) =
\sum_{i\in\tau} \sum_\sigma \left\{ \phi_i^*(\xi) X_i(\xi,\omega)
+ Y_i^*(\xi,\omega) \phi_i(\xi) \right\} ,
\end{equation}
where $\tau=n$ or $p$.
In this section,
we adopt the external field of $D_z^{E1}$ ($K=0$).
Using $\delta\rho(\omega)$ with $\omega=E+i\gamma/2$,
the strength function can be written as
\begin{eqnarray}
\frac{dB(E;D_z^{E1})}{dE} &=& -\frac{1}{\pi} {\rm Im} \int d\vec{r} \nonumber\\
   && \left(\frac{Ne}{A}z
      \delta\rho_p(\vec{r},\omega)
       -\frac{Ze}{A}z
      \delta\rho_n(\vec{r},\omega) \right) \nonumber\\
&=& \frac{2E\gamma}{\pi} \sum_n
\frac{ \tilde{E}_n | \bra{n} D_z^{E1} \ket{0} |^2 }
     {\left( E^2 - \tilde{E}_n^2 \right)^2 + E^2\gamma^2 } .
\end{eqnarray}
Substituting $\bra{n} D_z^{E1} \ket{0}$ by the following expression
\begin{equation}
\bra{n} D_z^{E1} \ket{0} = \int d\vec{r}
\frac{Ne}{A}z \bra{n} \hat{\rho}_p(\vec{r})\ket{0}
     -\frac{Ze}{A}z \bra{n} \hat{\rho}_n(\vec{r})\ket{0} ,
\end{equation}
we find
\begin{equation}
{\rm Im}\ \delta\rho_{n(p)}(\vec{r},\omega) \propto
\sum_n \frac{ \tilde{E}_n \bra{0} D_z^{E1} \ket{n}
                          \bra{n}\hat{\rho}_{n(p)}(\vec{r})\ket{0} }
     {\left( E^2 - \tilde{E}_n^2 \right)^2 + E^2\gamma^2 } ,
\end{equation}
where $\hat{\rho}_\tau(\vec{r})=\sum_{i\in \tau} \delta(\vec{r}-\vec{r}_i)$.
If there is a single state $\ket{n}$ that gives a dominant contribution
in the vicinity of $E=\tilde{E}_n$ (within the width of $\gamma$),
we have 
${\rm Im}\ \delta\rho(\omega)\propto \bra{n}\hat{\rho}\ket{0}$.

The left panels of Fig. \ref{drho} show
the calculated transition densities of neutrons and protons
separately, at $\omega=19.4 + 0.5 i$ MeV for
$^{16}$O and at $\omega=16.1 + 0.5 i$ MeV for the $K=0$ mode of $^{24}$Mg. 
These energies correspond to the peak energies of the GDR.
In the transition density of $^{16}$O, one can clearly 
see the isovector character.
In the top-right panel of Fig. \ref{drho}, we  show
$\partial \rho_0/\partial r$ for $^{16}$O,
where the $\rho_0(r)$ is the ground-state
density profile.
This corresponds to the transition density predicted by
the Tassie model for an irrotational and incompressible fluid.
As is discussed in the literatures \cite{RS80}, the GDR of light nuclei
is approximately represented by this classical fluid model.
On the other hand, the transition density of $^{24}$Mg
shows more complicated features.
We show, in the bottom-right panel of Fig. \ref{drho},
a contour plot of the transition density in the Tassie model,
$\delta\rho\propto
 \nabla \rho_0 \cdot \nabla rY_{10}(\hat{r})\sim \partial\rho_0/\partial z$.
Compared to the RPA transition density, there is a deviation, especially
in the nodal structure in the interior region.
This may suggest the importance of
coupling of the dipole mode to the octupole and higher-multipole modes.

\subsection{Heavy nuclei}
\label{sec:heavy_nuclei}

\begin{figure}[tb]
\begin{center}
\includegraphics[width=0.48\textwidth,keepaspectratio]{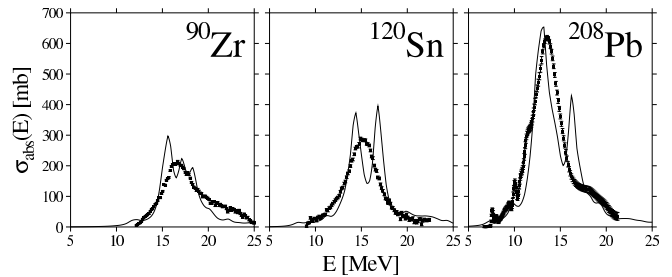}
\caption{{\small 
Calculated photoabsorption cross sections
for $^{90}$Zr, $^{120}$Sn, and $^{208}$Pb.
The calculation has been performed with $R_{\rm box}=15$ fm,
the SkM$^*$ parameter set, and $\gamma=1$ MeV.
The experimental data (symbols) are taken from
Refs. \cite{Lepretre71, Lepretre74, Beljaev91}.
}}
\label{fig.comp.exp.heavy}
\end{center}
\end{figure}

For heavier nuclei, the calculation better agrees with experiments.
As is well-known, in heavy spherical nuclei, a single energy-weighted
Lorentzian curve can fit the experimental data of 
the photoabsorption cross section very well \cite{BF75}.
We calculate photoabsorption cross sections 
in spherical nuclei $^{90}$Zr, $^{120}$Sn, and $^{208}$Pb within a box of 
radius $R_\mathrm{box}=15$ fm and compare them with experimental data
in Fig. \ref{fig.comp.exp.heavy}. 
The calculated GDR peak shows a splitting, however, this may be due to
the spurious effect coming from the box discretization.
Except for this splitting, the results agree well with the experimental data.
Table \ref{table.comp.exp} shows the calculated and the experimental
values of the GDR peak positions and the widths.
Both values are extracted from
the Lorentzian fit to the photoabsorption cross section.
According to Ref.~\cite{BF75},
the Lorentzian function of the form (\ref{dB/dw_gamma}),
\begin{equation}
\sigma_{\rm abs}(E)=\frac{\sigma_0}{1+(E^2-E_{\rm peak}^2)^2/(E^2\Gamma^2)} ,
\end{equation}
is used.
The GDR peak positions are well reproduced 
within an error of 400 keV.
The systematic deviation of the peak energies, seen in calculations for
light nuclei, cannot be observed for these heavy systems.
Therefore, we may conclude that the SkM$^*$ functional
reproduces peak energies of the $E1$ resonances in heavy nuclei.

The calculated widths are slightly smaller than the experiments,
by $300\sim 700$ keV.
Since we use $\gamma=1$ MeV in these calculations,
this indicates the spreading width of
$\Gamma^\downarrow=1.3\sim 1.7$ MeV.
The total damping width is about 4 MeV for these nuclei.
Thus, the spreading width is less than half of the total width.
This is rather surprising because,
for heavy nuclei, the spreading width was supposed to be a major
part of the total damping width \cite{Wam88,HW01}.
However, the fragmentation of the strength into non-collective 1p-1h
states (Landau damping), which can be described by the RPA theory,
is significant for these heavy systems.
Thus, the spreading width of $\Gamma^\downarrow\sim 1.5$ MeV is able to
reproduce a broadening of the experimental strength distribution.
This observation seems consistent with recent self-consistent
RPA calculations for spherical nuclei \cite{Ter06,Sil06}.

\begin{table}
\caption{
Calculated and experimental \cite{BF75} GDR peak energies and widths for
$^{90}$Zr, $^{120}$Sn, and $^{208}$Pb.
The FAM calculation was performed with the same parameter set
as Fig.~\ref{fig.comp.exp.heavy}.
}
\label{table.comp.exp}
\begin{center}
\begin{tabular}{|c|c|c|c|c|}
\hline
 & \multicolumn{2}{|c|}{Calculation} & \multicolumn{2}{|c|}{Experiment} \\ \cline{2-5}
Nucleus & $E_\mathrm{peak}$ [MeV] & $\Gamma$ [MeV] & $E_\mathrm{peak}$ [MeV] & $\Gamma$ [MeV] \\
\hline
$^{90}$Zr  & 16.37 & 3.85 & 16.74 & 4.16 \\
$^{120}$Sn & 15.22 & 4.52 & 15.40 & 4.89 \\
$^{208}$Pb & 13.26 & 3.47 & 13.63 & 3.94 \\
\hline
\end{tabular}
\end{center}
\end{table}

\section{Conclusions}
\label{sec:conclusion}

We have presented an implementation of the finite amplitude
method (FAM) to make a fully self-consistent RPA calculation
employing a realistic Skyrme energy functional. Although the
RPA is a well established theory and have been widely applied,
a fully self-consistent calculation with a realistic functional
is rather limited. This is mostly because of the complexity
to construct the residual two-body kernels for realistic
functionals. The FAM, which was proposed recently by the
present authors, makes it possible to achieve a fully
self-consistent RPA calculation without an explicit
construction of the residual two-body kernels. Instead, the
residual fields are evaluated by the finite difference,
employing a computational code for the static mean-field
Hamiltonian alone with a minor modification.

We implemented the FAM in the mixed representation where
the orbital indices are used for the occupied orbitals
while the three-dimensional Cartesian grid points are
used for the unoccupied orbitals. In this representation,
the linear response equation is a linear algebraic problem
with a sparse and non-hermitian matrix. We have tested several
solvers for the problem, and have found that the generalized
conjugate residual method works stably to obtain the solution.
We also examined the accuracy of the FAM employing a different
parameters of the finite difference, and have found that the
method works robust, namely, accurate results can be obtained
for a certain wide range of finite difference parameter.

We show results of the fully self-consistent RPA calculation
for the electric dipole responses of several nuclei of
spherical and deformed shapes. For light nuclei, we have found
a systematic underestimation of the average excitation energy,
irrespective of the energy functional employed. We further notice
the underestimation of the strength above the giant dipole
resonance. The discrepancy is significant only for light nuclei,
the average excitation energies of heavy nuclei are reasonably
described. For a deformed nucleus $^{24}$Mg, the calculation
shows a deformation splitting of the giant dipole resonance,
which well agrees with measurements. For spherical heavy nuclei,
we have found a substantial part of the width can be explained
within the RPA, leaving less than half of the total width for
a spreading width, in contrast to the previous studies which
reported that most of the width is attributed to the spreading
mechanism for heavy nuclei.

We are currently performing a systematic calculation of the
electric dipole responses for light to medium mass nuclei using
the present method and Skyrme energy functionals. In a forthcoming
paper, we will report a systematic analysis of the properties of
the giant dipole resonance, including the average excitation
energy, width, and the low-lying dipole mode around the threshold.

\section*{Acknowledgments}

This work is supported by Grant-in-Aid for Scientific Research on Innovative
Areas (No. 20105003) and by the Grant-in-Aid for Scientific Research(B)
(No. 21340073).
Computational resources were provided by the PACS-CS project and
the Joint Research Program
(07b-7, 08a-8, 08b-24) at Center for Computational Sciences,
University of Tsukuba, by the Large Scale Simulation Program
(Nos. 07-20 (FY2007), 08-14 (FY2008)) of High Energy Accelerator Research
Organization (KEK).
Computations were also performed on the RIKEN Super Combined
Cluster (RSCC).
We also thank the International Research Network for ``Exotic Femto Systems''
(EFES) of the Core-to-Core Programs of JSPS
and the UNEDF SciDAC collaboration.

\bibliography{myself,nuclear_physics,chemical_physics,current}

\end{document}